\documentclass[preprint,review,12pt]{elsarticle}

\usepackage{graphicx}

\usepackage{amssymb}
\usepackage{amsbsy}

\biboptions{numbers,comma,square,sort&compress}

\def\gtwid{\mathrel{\raise.3ex\hbox{$>$\kern-.75em\lower1ex\hbox{$\sim$}}}}
\def\ltwid{\mathrel{\raise.3ex\hbox{$<$\kern-.75em\lower1ex\hbox{$\sim$}}}}

\journal{Nuclear Physics B}

\begin{document}

\begin{frontmatter}



\title{Stripe Structures in the $t$-$t^\prime$-$J$ Model}


\author{D.J.~Scalapino}

\address{Physics Department, University of California, Santa Barbara, CA 93106-9530, USA}

\author{S.R.~White}

\address{Physics Department, University of California, Irvine, CA 92697, USA}

\begin{abstract}

Here, based upon density matrix renormalization group calculations, we discuss
the structure of the stripes found in the doped $t$-$t^\prime$-$J$ model and
the physics that underlies their formation.

\end{abstract}

\begin{keyword}

$t$-$t^\prime$-$J$ \sep stripes \sep DMRG


\end{keyword}

\end{frontmatter}


\section{Introduction}
\label{sec:1}

From the variety of experiments described in this review issue, it is
clear that stripes, or stripe-like fluctuations, appear as groundstates or
important low-energy configurations in the underdoped cuprates. Furthermore,
in the presence of impurities or lattice defects, the interplay between
randomness and the stripe-like correlations can lead to the electronic
inhomogeneity observed in these materials at low doping \cite{ref:1}. There have been
various theoretical models and explanations for the occurrence of these
stripes. Early on, Hartree-Fock calculations \cite{ref:2,ref:3,ref:4,ref:5} for the doped Hubbard
model found that one-dimensional domains of increased hole density forming
anti-phase N\'eel boundaries were present in the mean-field solutions. Here,
the stability of the stripe structure arose from the reduction in the kinetic
energy that the holes experienced in moving transverse to the stripes. The
stripes in this Hartree-Fock solution were insulating, characterized by a
filling of one hole per domain wall unit cell, while experiments on the
cuprates found a filling of half this \cite{ref:6}. An alternate view proposed
that stripes could arise from a competition between phase separation and
long-range Coulomb interactions \cite{ref:7}. In this ``frustrated phase separation" approach,
it was argued that lightly doped $t$-$J$ and Hubbard models would, in the absence
of a long-range Coulomb interaction, globally phase separate into uniform
hole-rich and undoped regions. While there are disagreements regarding whether
the simple two-dimensional $t$-$J$ or Hubbard models, with parameters in the
relevant physical regime, will in fact exhibit phase separation, if phase
separation occurs, the formation of stripes in this scenario depends upon the
suppression of the phase separation by the long-range Coulomb interaction. In
this case, the stripes can be metallic. In a third scenario, which we will discuss,
the formation of stripes in the 2D $t$-$J$ model is found to involve,
in addition to the reduction of kinetic energy of the holes arising from the
$\pi$-phase shifted antiferromagnetic stripe correlations, $d$-wave pairing
correlations leading to half-filled stripes \cite{ref:8,ref:9}.

In the following, we will review density matrix renormalization group \cite{ref:10} (DMRG)
results for the $t$-$J$ model. The Hamiltonian for the basic $t$-$J$ model is
\begin{equation}
  H=-t\sum_{\langle ij\rangle s}(c^\dagger_{is}c_{js}+{\rm h.c.})+J\sum_{\langle ij\rangle}
  \left({\bf S}_i\cdot{\bf S}_j-\frac{n_in_j}{4}\right)
	\label{eq:1}
\end{equation}
Here $\langle ij\rangle$ are nearest-neighbor sites, $s$ is a spin index and
doubly occupied sites are explicitly excluded from the Hilbert space. The
operator $c^\dagger_{is}$ creates an electron of spin $s$ on site $i$ and
${\bf S}_i=\frac{1}{2}c^\dagger_{is}{\boldsymbol\sigma}c_{is\prime}$ and
$n_i=c^\dagger_{i\uparrow}c_{i\uparrow}+c^\dagger_{i\downarrow}c_{i\downarrow}$ are the
electron spin moment and charge density operators at site $i$. The nearest-neighbor
hopping and exchange interaction are $t$ and $J$, respectively, and the average
site occupation $\langle n\rangle=1-x$ is set by the hole doping parameter $x$.
A next-near-neighbor hopping $t^\prime$ will also be considered. Except for
cases in which a proximity pairfield is applied, the particle number will be fixed.

\section{DMRG Results for the $t$-$J$ Model}
\label{sec:2}

\begin{figure}[!h]
\centerline{\includegraphics[width=0.5\textwidth]{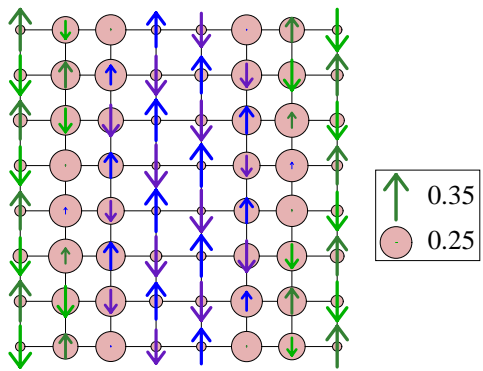}}
\caption{Hole density and spin moments on a center section of a 
$16 \times 8$ $t-J$ lattice with
$J/t = 0.35$ and an average hole density $x = 0.125$. The diameter of the circles is proportional to the
hole density $\langle n_i\rangle $ on the $i$th site and the length of the arrows is proportional to $\langle S^z_i\rangle $ according
to the scales shown. The arrows are color coded to show different antiferromagetic domains. This
structure depends on the boundary conditions as discussed in the text. (From \protect\cite{ref:8}.)}
\label{fig:1}
\end{figure}

Figure~\ref{fig:1} shows DMRG results for the central $8\times8$ section of a
$16\times8$ lattice which has periodic boundary conditions in the $y$-direction
and open end boundary conditions in the $x$-direction forming an open ended
cylinder. Here, $J/t=0.35$ and there are 16 holes corresponding to a doping
$x=1/8$. Along the left and right ends of the cylinder a small staggered
magnetic field of magnitude $h=0.1t$ was applied. This field breaks the spin
symmetry and the open ends break the translational symmetry in the $x$-direction,
giving rise to nonhomogeneous finite values of the hole occupation $\langle n_i\rangle$ and
spin $\langle S^z_i\rangle$. The basic pattern consists
of domain walls containing excess hole density separated by $\pi$-phase shifted
antiferromagnetic regions. There are four holes per domain wall, and as we will
discuss this linear filling of 0.5 holes per unit length of the domain wall is
the preferred filling. Here the boundary conditions only act to pin the stripes
and calculations with different boundary conditions provide evidence that
the basic stripe pattern is intrinsic.

\begin{figure}[!h]
\centerline{\includegraphics[width=0.5\textwidth]{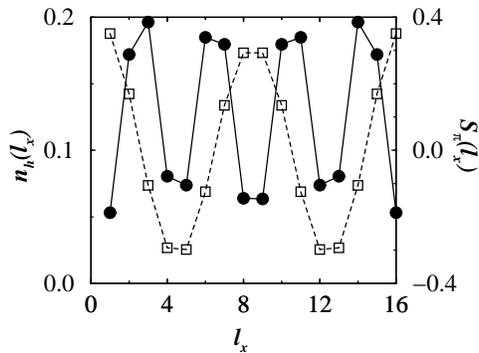}}
\caption{Average hole density $n_h(\ell_x)$ (solid circles) and spin structure
function $S_{\pi}(\ell_x)$ (open squares) versus $\ell_x$ 
for the $16\times 8$ system of Fig.~\protect\ref{fig:1}. \protect\cite{ref:11}}
\label{fig:2}
\end{figure}
A more detailed look at the domain wall structure is shown in Fig.~\ref{fig:2}.
Here the $\ell_x$ dependence of the average hole density
\begin{equation}
  n_h(\ell_x)=\frac{1}{L_y}\sum^{L_y}_{\ell_y=1}
  \left(1-\left\langle c^\dagger_{\ell_x\ell_{y\uparrow}}c_{\ell_x\ell_{y\uparrow}}+
  c^\dagger_{\ell_x\ell_{y\downarrow}}c_{\ell_x\ell_{y\downarrow}}\right\rangle\right)
  \label{eq:2}
\end{equation}
and the spin structure
\begin{equation}
  S_\pi(\ell_x)=\frac{1}{L_y}\sum^{L_y}_{\ell_y=1}(-1)^{\ell_x+\ell_y}
  \left\langle S^z\left(\ell_x,\ell_y\right)\right\rangle
	\label{eq:3}
\end{equation}
are shown as a function of the $x$-coordinate $\ell_x$. The
charge structure of these domain walls is evident as is the $\pi$-phase
shifted antiferromagnetic regions separating them.

\begin{figure}[!h]
\centerline{\includegraphics[trim=0mm 0mm 0mm 1mm, clip=true, width=0.5\textwidth]{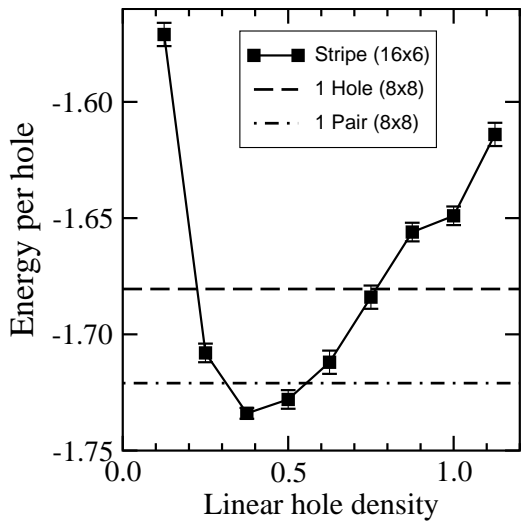}}
\caption{Energy per hole for a $16\times6$ stripe versus 
the linear density of holes $\rho_\ell$. 
The horizontal lines mark the energy per 
hole of one hole (dashed line) and two holes (dot-dashed line) 
placed on an open $8\times8$ lattice with a staggered field  of magnitude 0.1 on all
four sides \protect\cite{ref:12}.} \label{fig:3}
\end{figure}
In Fig.~\ref{fig:3}, we show the energy per hole as a function of the linear
hole density $\rho_\ell$ measured on a $16\times6$ system with open end
boundary conditions and with a staggered $\pi$-phase shifted magnetic field
of magnitude $h=0.1$ applied to the top and bottom rows of sites. There is
a minimum for a linear density $\rho_\ell\sim0.5$. The dashed line on this plot
shows the energy of one hole on an $8\times8$ lattice with a staggered field of
magnitude 0.1 on all four sides while the dash-dot line is the energy per hole
of two holes on this lattice. The fact that the energy per hole of two holes is
lower than that of a single hole indicates that two holes pair in agreement with
exact diagonalization studies. However, we note that the energy per hole
associated with the domain wall for $\rho_\ell\sim0.5$ is even lower suggesting
that pairs condense into domain walls even at low doping. The interaction between
domain walls was also studied and was found to be repulsive.

Thus for $x\ltwid0.125$,
$\rho_\ell\sim0.5$ (1,0) domain walls are favored and their spacing $d$ is
equal to $(2x)^{-1}$. For $x\gtwid0.17$, the DMRG calculations find
$\rho_\ell\simeq1$ domain walls begin to form. At this filling the energy per
hole for the (1,0) and (1,1) walls, as well as the pairs, are nearly degenerate
leading to large fluctuations in the configurations \cite{ref:12}.

In addition to the charge and spin structure of the domains, they also exhibit
an underlying pairfield structure which can be described in terms of the
near neighbor pairfield operators
\begin{equation}
  \Delta^\dagger_x(\ell_x,\ell_y)=\frac{1}{\sqrt2}\left(c^\dagger_\uparrow\left(\ell_x+1,
  \ell_y\right)c^\dagger_\downarrow\left(\ell_x,\ell_y\right)-c^\dagger_\downarrow
  \left(\ell_x+1,\ell_y\right)c^\dagger_\uparrow\left(\ell_x,\ell_y\right)\right)
	\label{eq:4}
\end{equation}
and
\begin{equation}
  \Delta^\dagger_y(\ell_x,\ell_y)=\frac{1}{\sqrt2}\left(c^\dagger_\uparrow\left(\ell_x,\ell_y+1\right)
  c^\dagger_\downarrow\left(\ell_x,\ell_y\right)-c^\dagger_\downarrow\left(\ell_x,\ell_y+1\right)
  c^\dagger_\uparrow\left(\ell_x,\ell_y\right)\right)
	\label{eq:5}
\end{equation}
The first term creates a singlet pair on the $(\ell_x+1,\ell_y)-(\ell_x,\ell_y)$ link
of the lattice and the second term creates a pair on a corresponding $y$-link. Since the stripes have
broken the $C_4$ rotational symmetry of the lattice, the pairfield-pairfield
correlations for $x$-link pairs differ from those of the $y$-link pairs.
However, the $d$-wave-like nature of the pairing correlations is evident in
the fact that $\left\langle\Delta_y\left(\ell^\prime_x,\ell^\prime_y\right)
\Delta^\dagger_x\left(\ell_x,\ell_y\right)\right\rangle$ is found to be negative\cite{ref:9}.

Based on the behavior of the two-leg $t$-$J$ ladder \cite{ref:13a,ref:13b}, it is natural to expect that
the stripes will exhibit pairfield correlations. For the two-leg ladder, one
knows that these correlations have a power law decay and indeed if a sufficient
number of states are kept in the DMRG calculations, a power law decay is observed.
However, the off diagonal nature of the pairing correlations makes this one of
the more challenging calculations.

\begin{figure}[!h]
\centerline{\includegraphics[width=0.5\textwidth]{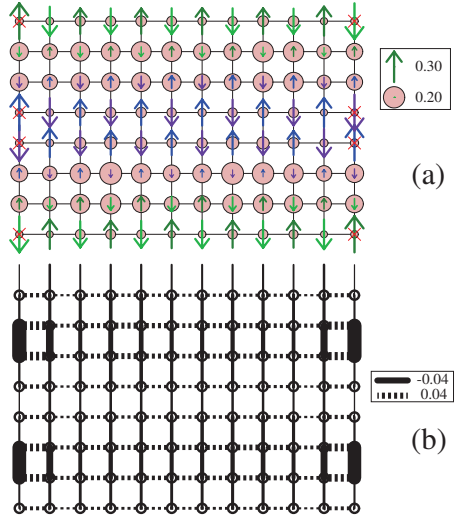}}
\caption{(a) A $12\times18$ $t$-$J$ lattice with cylindrical boundary
conditions in the $y$-direction with $J_x=0.55$, $J_y=0.45$ such that the
stripes run along the $x$-axis. Here we have applied a magnetic field $h=\pm0.2$ 
on the sites with a red X, and a pair field $\Delta_0=1.0$ to the edge links
without X's.  A chemical potential $\mu=1.23$ was used to give a doping of
$x=0.127$. (b) The pair field strength
$\langle\Delta^+_\alpha(\ell_x,\ell_y)+\Delta_\alpha(\ell_xk,\ell_y)\rangle$
on the $\alpha=x$ and $y$ links for the system shown in (a) \protect\cite{ref:9}.
}
\label{fig:4}
\end{figure}

In addition, for the striped systems of interest, there are further difficulties.
Consider the $16\times8$ lattice shown in Fig.~\ref{fig:1}, where the boundary conditions
are such that the stripes run around the circumference of the cylinder with each
stripe on average containing 4 holes corresponding to a linear hole density
$\rho=0.5$. Fluctuations in which a pair of holes is exchanged between two of
the stripes leaving one stripe with 2 holes and another with 6 holes lead to
linear hole densities of 0.25 and 0.75, respectively in the two stripes. From
Fig.~\ref{fig:3}, one can see that such large deviations of the linear hole density are
energetically costly so that pair number fluctuations are suppressed leading in
turn to the suppression of the pairfield correlations. In order to study
longer stripes which can more readily support density fluctuations, we have
considered systems which have slightly anisotropic exchange interactions
$(J_x=0.55,\ J_y=0.45)$\cite{ref:9}. As shown in Fig.~\ref{fig:4}a, this anisotropy favors
stripes that are oriented along the long $x$-axis of a $12\times8$ cylinder.
Here, in addition to the magnetic fields, a proximity pairfield coupling
$\Delta_0(\ell_x,\ell_y)\left(\Delta^\dagger_y\left(\ell_x,\ell_y\right)+\Delta_y
\left(\ell_x,\ell_y\right)\right)$ was applied to rungs at both ends of the
cylinder. In the presence of this proximity pairfield, the total number of
electrons is only conserved modulo two, and a chemical potential $\mu$ was
introduced to control the average number of holes.

Figure~\ref{fig:4}b shows the expectation value of the proximity induced
pairfield $\left\langle\Delta^\dagger_\alpha\left(\ell_x,\ell_y\right)+\Delta_\alpha
\left(\ell_x,\ell_y\right)\right\rangle$ for $\alpha=x$ and $y$ with a coupling
strength $\Delta_0(\ell_x,\ell_y)=1$ for the four thickest end links and 0.5 for
the four vertical links adjacent to them. The $d$-wave-like sign changes of the
induced pairfield are clearly seen in Fig.~\ref{fig:4}b. If the proximity
pairfield is applied to the end rungs of only one of the domains, the structure
of the induced pairfield pattern remains similar, but the amplitude of the
induced pairfield on the domain that is not coupled to the proximity pairfield
becomes considerably weaker. This again reflects the fact that
it is energetically unfavorable for the number of hole pairs in a domain of
this length which is not coupled to an external pair reservoir to have
significant hole number fluctuations.

\begin{figure}[!h]
\centerline{\includegraphics[width=0.5\textwidth]{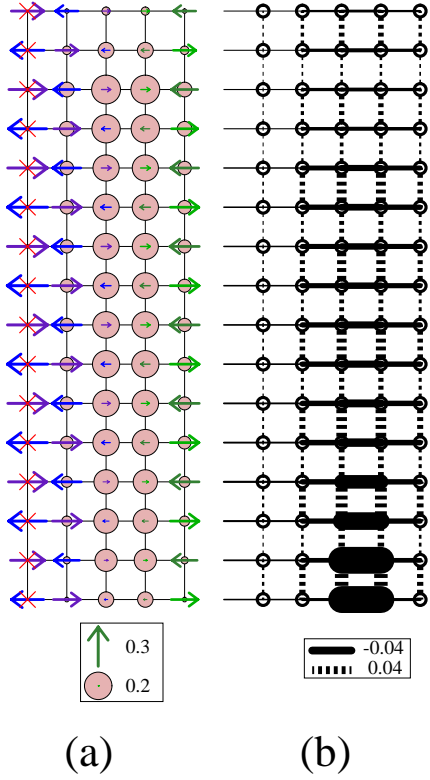}}
\caption{A single forced stripe on a $5\times16$ lattice with cylindrical
boundary conditions. Here, $J=0.5$ and $t'=0.2$. On the $x=5$ leg, a staggered
field $(-1)^yh$ with $h=0.05$ was applied, along with a chemical potential of
2.0. On the other 4 legs, a chemical potential $\mu$ was applied to vary the
doping. In the case shown $\mu=1.41$, yielding a doping of $x=0.106$, which
corresponds to a linear doping of $0.53$. In each case a strong pair field was
applied to four rungs ($y=1$ to 4, connecting $x=3$ and $x=4$), visible as the
four thickest links in (b). The magnitude of the applied field on the four
rungs was 1.0, 1.0, 0.5, 0.25. (a) The hole density $\langle 1-n_i\rangle$ and
spin density $\langle S^z_i\rangle$. (b) The measured pair field
$\left\langle\Delta^\dagger_\alpha\left(\ell_x,\ell_y\right)+\Delta_\alpha
\left(\ell_x,\ell_y\right)\right\rangle$ \protect\cite{ref:9}.
}
\label{fig:5}
\end{figure}

\begin{figure}[!h]
\centerline{\includegraphics[width=0.5\textwidth]{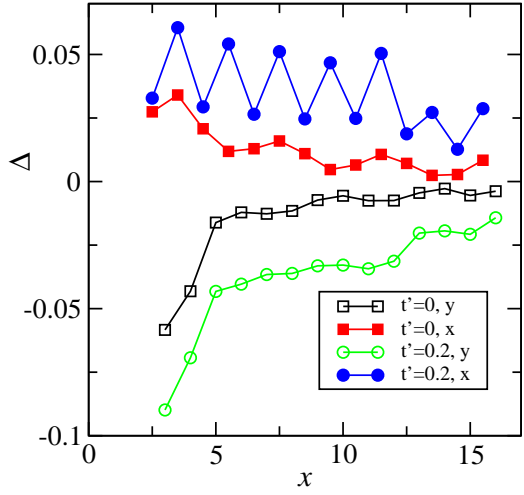}}
\caption{The 3--4 rung 
$\left\langle\Delta^\dagger_x\left(3,\ell_y\right)+\Delta_x
\left(3,\ell_y\right)\right\rangle$  
and the third leg
$\left\langle\Delta^\dagger_y\left(3,\ell_y\right)+\Delta_y
\left(3,\ell_y\right)\right\rangle$ induced pairfields versus $\ell_y$ for
the $5\times16$ lattice of Fig.~\protect\ref{fig:5}.
}
\label{fig:6}
\end{figure}

\begin{figure}[h!]
\centerline{\includegraphics[trim=0mm 0mm 0mm 1mm, clip=true, width=0.5\textwidth]{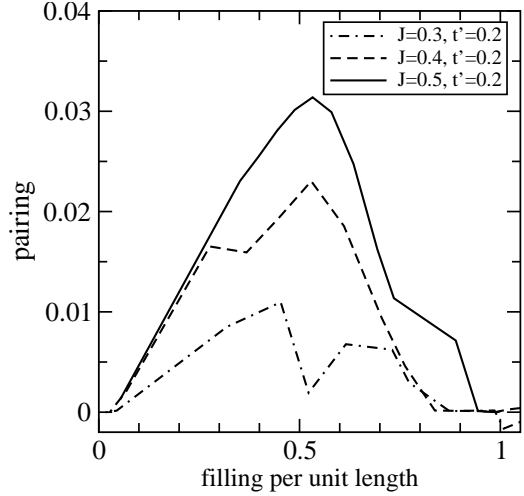}}
\caption{The measured pair field at the $y=12$, $x=3$--4 rung versus the
number of holes per unit length \protect\cite{ref:9}.}
\label{fig:7}
\end{figure}

As seen in Fig.~\ref{fig:4}b, the proximity pairfield induces a pairfield
which spreads out over the lattice but decays in strength as one moves away
from the source. In order to look in more detail at this induced pairfield,
it is useful to separate questions dealing with pairing on a stripe and
pairing between stripes. For the first question, we have studied a single
stripe on the $5\times16$ lattice with cylindrical periodic boundary
conditions shown in Fig.~\ref{fig:5}a. In this case, the periodic boundary
conditions in the cylindrical direction give rise to one stripe, similar to
those shown in Fig.~\ref{fig:1}. Then, applying a proximity pairfield to only
one end of the stripe one obtains the induced pairfield shown in Fig.~\ref{fig:5}b.
The strength of the induced pairfield depends upon $J$, the next near neighbor
hopping $t^\prime$ and the doping $x$. As previously found in various numerical
calculations \cite{ref:14,ref:15}, a positive value of $t^\prime$ favors pairing while a
negative value suppresses it. Figure~\ref{fig:6} shows the rung and leg induced
pairfields
$\left\langle\Delta^\dagger_x\left(3,\ell_y\right)+\Delta_x
\left(3,\ell_y\right)\right\rangle$  
and
$\left\langle\Delta^\dagger_y\left(3,\ell_y\right)+\Delta_y
\left(3,\ell_y\right)\right\rangle$  versus $\ell_y$
for a linear filling $\rho_\ell=0.5$ for $J=0.5$ and $t^\prime=0$
and 0.2. The $d$-wave-like character of the pairfield is evident. The
magnitude of the induced pairfield on the 12th $x=3$--4 rung is shown in
Fig.~\ref{fig:7} versus the linear hole doping $\rho_\ell$. When the pairing is
strongest, the response peaks for $\rho_\ell\sim0.5$ holes per unit length.

The second question involving the nature of the pairing between stripes was
studied using a $6\times12$ lattice with open boundary
\begin{figure}[!h]
\centerline{\includegraphics[width=0.5\textwidth]{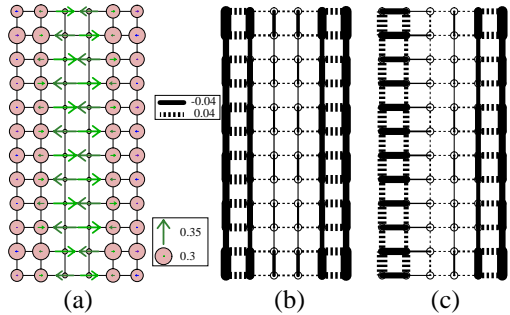}}
\caption{(a) Hole and spin densities for a $6\times12$ lattice with $J=0.5$
and $t'=0$, and with open boundary conditions in both directions with fields
applied to force two stripes.  The center two legs have an applied staggered
field $(-1)^yh$ with $h=0.05$, along with a chemical potential of 1.3. The
outer four legs have a chemical potential of 0.9, leading to a doping of
$x=0.1981$, corresponding to a linear doping per stripe of 0.594. The outermost
two legs have applied pair fields of 0.5 on each vertical rung. In (a), the
pair fields are applied in phase. (b) Measured pair fields for the in-phase case.
(c) Measured pair fields when the applied fields have opposite sign \protect\cite{ref:9}.
}
\label{fig:8}
\end{figure}
conditions in both the $x$ and $y$ directions. In this case, external magnetic
and chemical potential fields were used to force two stripes to form as
shown in Fig.~\ref{fig:8}a. Then the pairfield response was studied when an
external proximity pairfield was applied to the vertical links on the outermost
($x=1$ and 6) legs. In Fig.~\ref{fig:8}b, the applied proximity pairfields were
in phase and in case Fig.~\ref{fig:8}c they were out of phase. For $t^\prime$
negative the DMRG calculations found that energy difference between the
relative pair phase of the two domain walls was negligible, consistent with the
weak overall pairfield response found for $t^\prime<0$. For $t^\prime=0.0$ and
0.2, the energy to form an antiphase $d$-wave domain wall was found to be small
but positive \cite{ref:9}, of order 5\% of the energy difference per hole
between one hole and a pair shown in Fig.~\ref{fig:5} for $t'=0.2$. A similar
conclusion was reached in renormalized mean field theory calculations \cite{ref:16}.
Nevertheless, the difference in energy between the in phase and antiphase
$d$-wave pairfield states is small and there have been various suggestions
\cite{ref:17}, approximations \cite{ref:18}, and model calculations \cite{ref:19,Loder}
which find that the antiphase pairing state can have the lowest energy.

The picture that emerges from these DMRG calculations for the $t$-$J$ model is
that for dopings $x\ltwid0.125$, (1,0) domain walls are favored. These walls
have a linear hole filling $\rho_\ell=0.5$ and repel each other, leading to
a domain wall spacing $d=(2x)^{-1}$. The structure of the domain wall consists
of a stripe-like region of excess holes several lattice spacings in width
surrounded by antiferromagnetic regions with fewer holes which are $\pi$-phase
shifted across the stripe. The domain walls also have short range $d$-wave-like
pairfield correlations. A proximity pairfield applied to one domain creates a
$d$-wave-like response which extends in phase over a neighboring domain, but
appears within the present calculation to be short range. Depending upon the
boundary conditions, it is also possible to have site-centered domain walls.
The minimum energy filling remains $\rho_\ell=0.5$ and there are similar
$d$-wave pairfield correlations. In general, the stripes tend to occupy more
than one column or row so that there is not a sharp distinction between site- and bond-centered
stripes, and intermediate centerings are possible.

\section{Physical mechanism}
\label{sec:3}

In this section, we seek to determine the nature of the physics that underlies
the properties of the domain walls which have been described in Sec.~\ref{sec:2}.

To begin, there is of course the question of phase separation. Emery \cite{ref:21}
et al.\ originally argued that the 2D $t$-$J$ model phase separates for all $J/t$
interaction strengths close to half filling. Hellberg and Manousakis \cite{refHM},
using a Green's function Monte Carlo technique, found results that supported this.
However, Calandra et al.\ \cite{refCalandra}, who also used Green's function
Monte Carlo, concluded that phase separation only occured if $J/t>0.4$ and
Putikka \cite{refPutikka} et al., using a high temperature series expansion,
concluded that phase separation for dopings near half-filling required $J/t>1.2$.

The DMRG is known to be most effective for one-dimensional systems and ladder
systems of limited width. Therefore, we have studied the phase separation
question by examining a sequence of ladders which have an increasing number of
legs. For these $n$-leg ladders with $n$ up to 6, the DMRG results provide
reliable results for the $\langle n\rangle-J/t$ phase separation boundary.

\begin{figure}[h!]
\centerline{\includegraphics[width=0.5\textwidth]{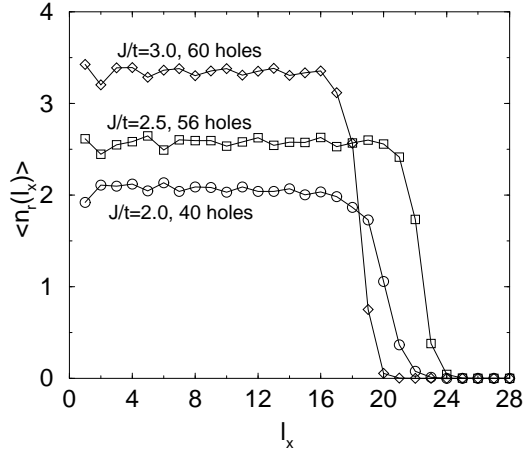}}
\caption{Density of holes on the rungs of a $4\times28$ system for three different values of $J/t$, all
of which allow phase separation \protect\cite{ref:20}.}
\label{fig:9}
\end{figure}

\begin{figure}[!h]
\centerline{\includegraphics[width=0.5\textwidth]{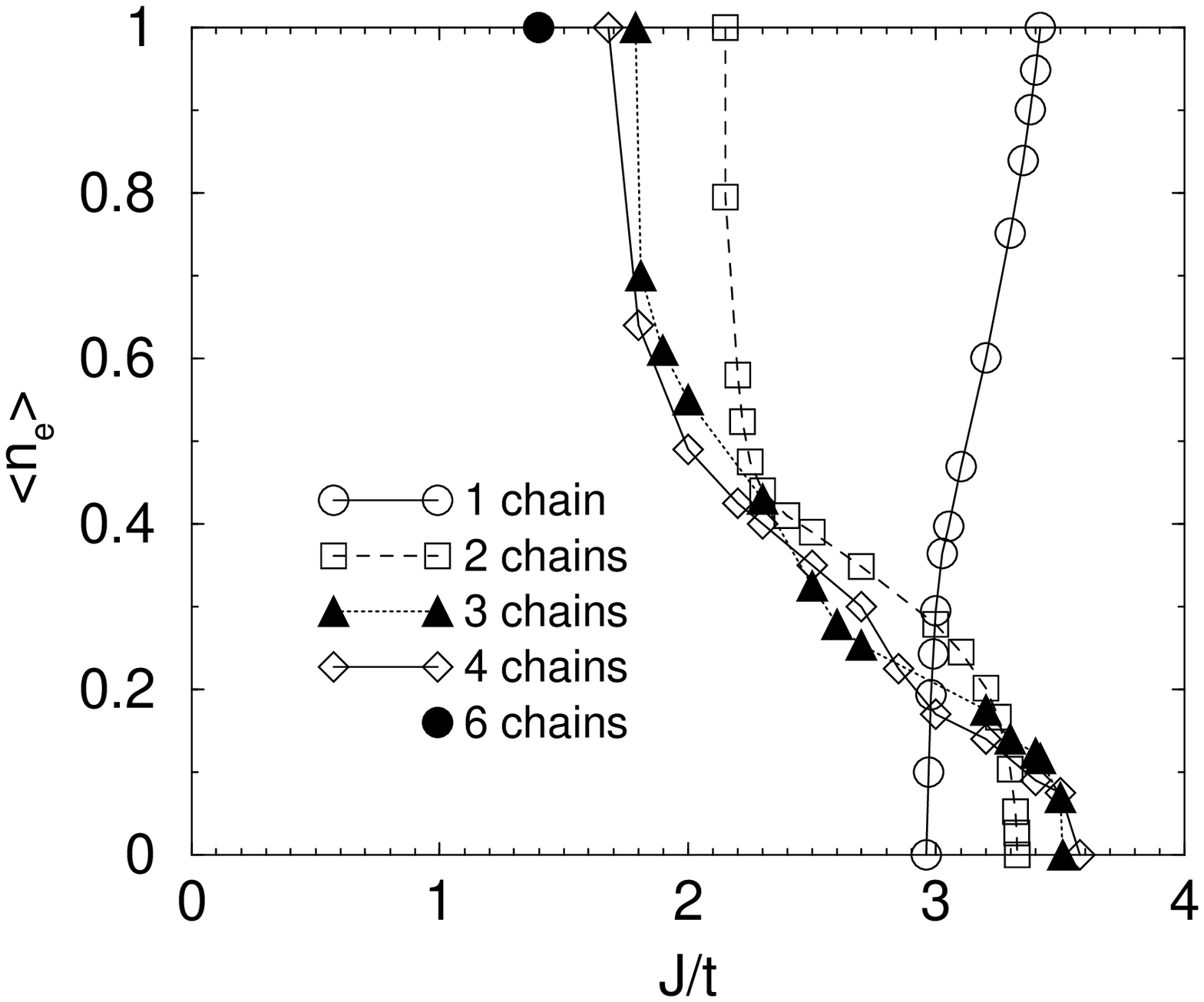}
\includegraphics[width=0.5\textwidth]{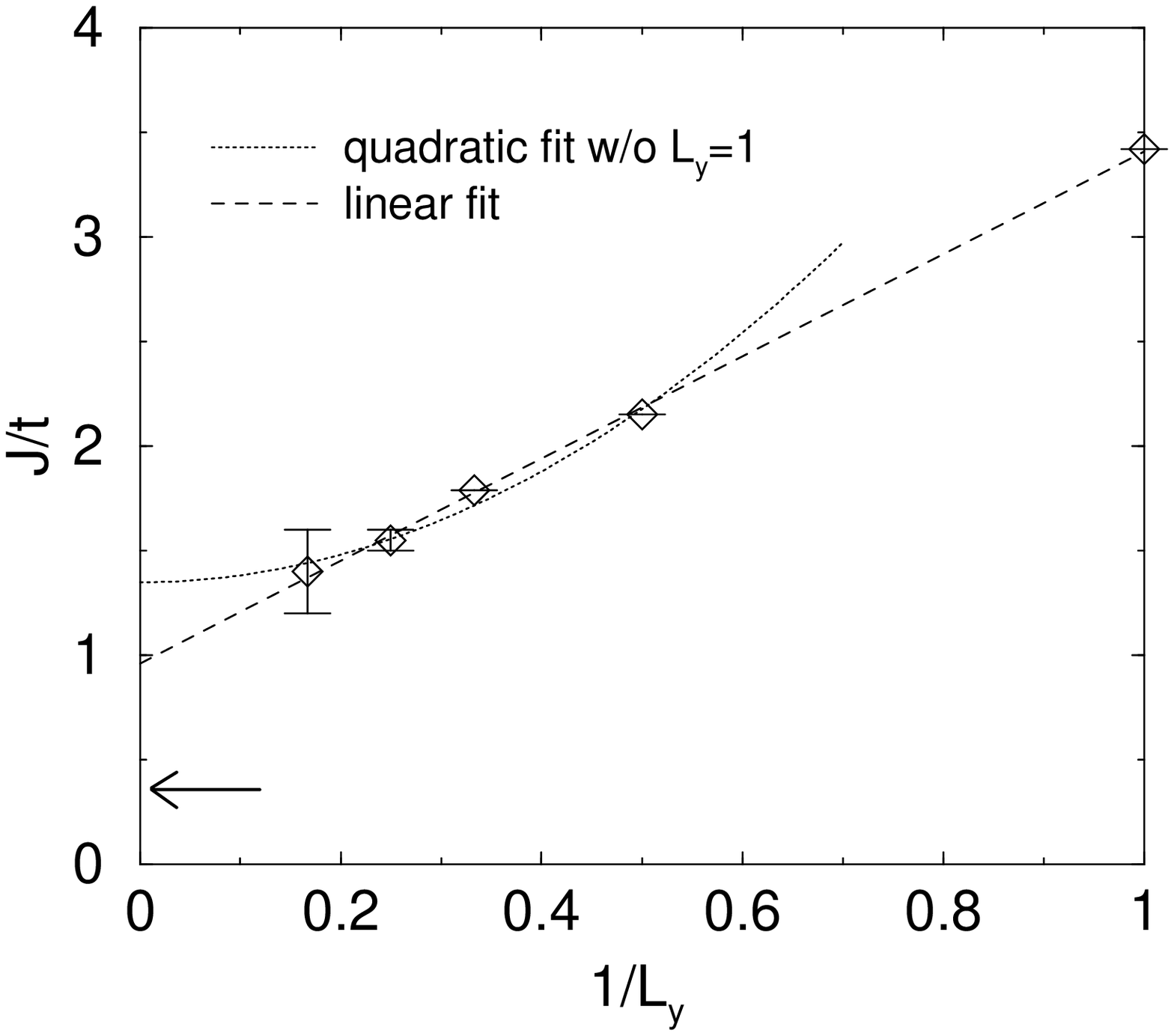}}
\caption{Left panel: Boundary to phase separated region where $\langle n_e \rangle$ is
the total electron density of the system. Phase separation is realized to
the right of the curves. Right panel: Phase separation boundary at low hole doping as a function of the
inverse number of legs. The arrow indicates $J/t \approx 0.35$, a typical
value of $J/t$ for the cuprates \protect\cite{ref:20}.}
\label{fig:10}
\end{figure}

In a $t$-$J$ ladder with two or more legs, the phase-separated system consists
of one region containing holes and one region without any holes. An example
of this for a 4-leg ladder is shown in Fig.~\ref{fig:9}.
For a given value of $J/t$ and hole doping $x$, the average number of holes
per site $\langle n_h\rangle$ in the hole-rich region determines a point on the
phase boundary between the phase separated and uniform systems. Some DMRG
results for the phase separation boundary showing $\langle n_e\rangle=1-\langle n_h\rangle$
versus $J/t$ for ladders with up to six legs are given in Fig.~\ref{fig:10}a.
In the absence of an exact functional form for the convergence to 2D, we have
plotted linear and quadratic fits to the inverse number of legs in Fig.~\ref{fig:10}b.
It appears from these results that the 2D system is away from the phase
separation boundary for physically relevant $J/t$ values and we believe that
the ground state of the 2D $t$-$J$ model at small $J/t$, close to half filling,
will be striped and not phase separated. For the present discussion, the important
point is that calculations on lattices we have discussed show no evidence of phase
separation in the $J/t$ parameter regimes of interest.

Next, there are the spin correlations whose $\pi$-phase shift across the
stripe are known to play an essential role in the Hartree-Fock solutions \cite{ref:2,ref:3,ref:4,ref:5}.
Here it is useful to introduce an operator
\begin{equation}
  {\mathbb P}(\ell_1,\ell_2,\cdots\ell_m)=\prod^m_{i=1}p(\ell_i)
\label{eq:6}
\end{equation}
with $p(\ell)=(1-n_{\ell\uparrow})(1-n_{\ell\downarrow})$ the hole projection
operator for the $\ell$th lattice site.
The operator ${\mathbb P}$ projects out a part of the groundstate that has $m$ holes
located at sites $\ell_1,\ell_2,\cdots\ell_m$. One can then separate the
wavefunction $|\psi\rangle$ into parts which have specific hole locations
\begin{equation}
  |\psi\rangle=\sum_{\{\ell_i\}}{\mathbb P}(\ell_i)|\psi\rangle=
	\sum_{\{\ell_i\}}a(\ell_i)|\psi_{\{\ell\}}\rangle
\label{eq:7}
\end{equation}
Here, $|\psi_\ell\rangle$ is a normalized wave function with holes at specific
sites and $a(\ell_i)> 0$. Using the DMRG, one can study $|\psi_{\{\ell\}}\rangle$
directly and then from it evaluate various expectation values \cite{ref:22}. We will focus
on the expectation value of ${\bf S}_i\cdot{\bf S}_j$. This measurement gives
one a ``snapshot" of the $\langle{\bf S}_i\cdot{\bf S}_j\rangle$ configuration
associated with a given hole configuration. We will use the term
``antiferromagnetic bond" or just ``bond" to indicate that
$\langle{\bf S}_i\cdot{\bf S}_j\rangle<0$. If this expectation value is close
to -0.75 for two sites $i$ and $j$, one would say that there is a ``singlet bond"
connecting $i$ and $j$, even if there is no term in the Hamiltonian directly
coupling $i$ and $j$.

\begin{figure}[!h]
\centerline{\includegraphics[width=0.5\textwidth]{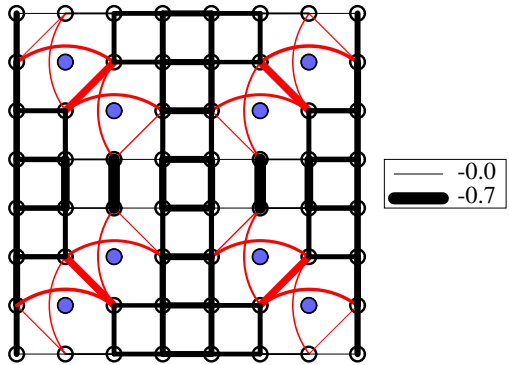}}
\caption{For an $8\times8$ system with
$x = 0.125$ whose hole density and spin patterns are similar to that shown in Fig.~\ref{fig:1}, the blue
dots show the most probable configuration of all the holes and the 
strength of the exchange bonds
$|\langle {\bf S}_i\cdot{\bf S}_j \rangle |$ between two sites is denoted by the thickness of the line connecting the sites. Anomalous
antiferromagnetic correlations across holes are colored red. Only correlations where 
$\langle {\bf S}_i\cdot{\bf S}_j\rangle < 0$
are shown \protect\cite{ref:8}.}
\label{fig:11}
\end{figure}

Using this projection technique, the spin configuration associated with the
domain structure of Fig.~\ref{fig:1} is shown in Fig.~\ref{fig:11}.
The blue circles show the most probable configuration of the eight holes on the
$8\times8$ section shown in Fig.~\ref{fig:1}. The thickness of the lines
connecting various sites denotes the strength of the exchange field
$\langle{\bf S}_i\cdot{\bf S}_j\rangle$ for this configuration of the holes.

In Fig.~\ref{fig:11} one sees that exchange bonds form across the mobile holes.
This reflects the local competition between the kinetic and exchange energies.
The domain wall allows most of the exchange bonds to form in a way which
cooperates, rather than competes, with the background spin configuration. In
particular, the domain wall forms to support $\pi$ phase-shifted
antiferromagnetic regions on either side to reduce the disturbance of the
exchange interactions and to lower the transverse kinetic energy of the holes.
This is the basic physics that underlies the Hartree-Fock results.

\begin{figure}[!h]
\centerline{\includegraphics[width=0.5\textwidth]{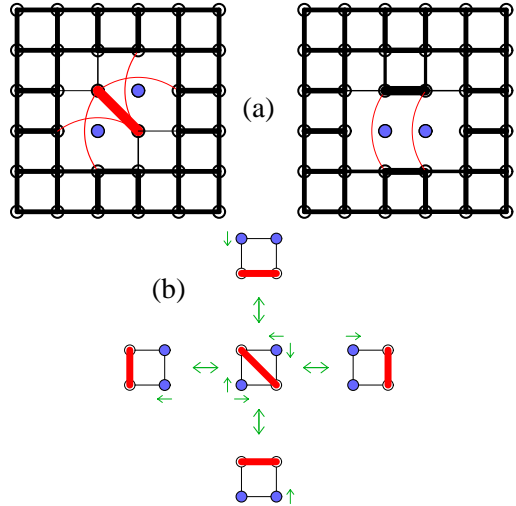}}
\caption{(a) The two most probable hole configurations and associated local exchange
field $\langle{\bf S}_i\cdot{\bf S}_j\rangle$ correlations in the two hole ground
state of the system shown in Fig.~\protect\ref{fig:11}. The scale and conventions
are the same as in Fig.~\protect\ref{fig:11}, but only the central $6\times6$
region is shown. (b) Schematic illustration showing the one-electron hops
from the diagonal configurations which lead to the low energy configurations
and contribute to a lowering of the kinetic energy \protect\cite{ref:8}.}
\label{fig:12}
\end{figure}

However, as we have seen, the domain walls are also characterized by $d$-wave
pairfield correlations. To see how these correlations are reflected in
Fig.~\ref{fig:11}, we examine this projection for a single pair. Two of the
``most probable" hole and spin configurations associated with the groundstate
of a pair of holes on an $8\times8$ lattice with $J/t=0.35$ and a staggered AF
($h=0.1$) boundary field are illustrated in Fig.~\ref{fig:12}.
Here, as before, the holes are shown as blue dots and the thickness of the
solid lines denote the singlet bond strength between sites. In the right panel,
the holes sit on adjacent sites and this configuration corresponds to what one
would expect when the exchange $J$ is large compared with the hopping $t$. In
this case, the holes occupy neighboring sites to reduce the number of broken
exchange interactions. However, in the physical regime with an exchange $J$
that is small compared with the hopping $t$, the dynamics is important and the
configuration shown in the left panel of Fig.~\ref{fig:12} is more probable.

This bound pair of holes is characterized by a $2\times2$ core region and to
better understand it we consider the $2\times2$ lattice shown in Fig.~\ref{fig:12}b.
Using the near neighbor singlet pair operator between sites $i$ and $j$,
\begin{equation}
  \Delta^\dagger_{ij}=\frac{1}{\sqrt2}(c^\dagger_{i\uparrow}c^\dagger_{j\downarrow}-c^\dagger_{i\downarrow}c^\dagger_{j\uparrow})
\label{eq:8}
\end{equation}
one can write the groundstate of the undoped half-filled system as
\begin{equation}
  |\psi\rangle_0=N_0[\Delta^\dagger_{14}\Delta^\dagger_{23}-\Delta^\dagger_{12}\Delta^\dagger_{34}]|0\rangle
\label{eq:9}
\end{equation}
with $|0\rangle$ the vacuum. The groundstate of the two-hole system is
\begin{equation}
  |\psi\rangle_2=N_2\left[a\left(\Delta^\dagger_{12}+\Delta^\dagger_{23}+\Delta^\dagger_{34}+\Delta^\dagger_{14}\right)
	+b\left(\Delta^\dagger_{13}+\Delta^\dagger_{24}\right)\right]|0\rangle
\label{eq:10}
\end{equation}
with $a=1$ and $b=\left(2+\left(\frac{J}{4t}\right)^2\right)^{1/2}-\left(\frac{J}{4t}\right)$.
Then in the doped, two-hole state $|\psi\rangle_2$, the ratio of the
near-neighbor singlet to the diagonal singlet amplitude is
\begin{equation}
  \frac{a}{b}=\frac{1}{\left[2+\left(\frac{J}{4t}\right)^2\right]^{1/2}-\frac{J}{4t}}
\label{eq:11}
\end{equation}
For $J/t=2$, this ratio is unity. For $J/t<2$, the diagonal amplitude is larger
than the edge amplitude. This is reflected in the $t$-$J$ results previously
discussed, where the hole-hole correlations were found to be larger for
next-nearest-neighbor diagonal sites than for nearest-neighbor sites.

The groundstate, Eq.~(\ref{eq:9}), of the undoped $2\times2$ system
transforms as $d_{x^2-y^2}$, while the two-hole state, Eq.~(\ref{eq:10}),
transforms as an $s$-wave. Thus, the hole-pair creation operator that connects
$|\psi\rangle_0$ to $|\psi\rangle_2$ must transform as $d_{x^2-y^2}$ 
\cite{ref:23,ref:24}.
A simple nearest-neighbor operator of this form is
\begin{equation}
  \Delta=\Delta_{14}-\Delta_{12}+\Delta_{23}-\Delta_{34}.
\label{eq:12}
\end{equation}
Applying this to the undoped groundstate $|\psi\rangle_0$ given by Eq.~(\ref{eq:9}),
one finds that
\begin{equation}
  \Delta|\psi\rangle_0=-2N_0[\Delta^\dagger_{12}+\Delta^\dagger_{23}+\Delta^\dagger_{34}+\Delta^\dagger_{14}]|0\rangle,
\label{eq:13}
\end{equation}
which clearly has a nonzero overlap with the two-hole  groundstate $|\psi\rangle_2$.

A $d_{x^2-y^2}$ hole pair creation operator, generalized to include holes on
next-nearest-neighbor diagonal sites, has been discussed by Poilblanc \cite{ref:25}.
One can expand a generalized hole-pair creation operator in terms of operators
which create a pair of holes on sites separated by a distance $R$. For our
$2\times2$ cluster, this involves
\begin{equation}
  \Delta_{d_{x^2-y^2}}=\sum_{R}\Delta_R,
\label{eq:14}
\end{equation}
with $R=1$ and $R=\sqrt2$. The nearest-neighbor operator $\Delta_1$ is just the
operator given in Eq.~(\ref{eq:12}) and a next-nearest-neighbor term possessing
$d_{x^2-y^2}$ symmetry \cite{ref:25} is
\begin{equation}
  \Delta_{\sqrt2}=({\bf S}_1-{\bf S}_3)\cdot{\bf T}_{24}-({\bf S}_2-{\bf S}_4)\cdot{\bf T}_{31},
\label{eq:15}
\end{equation}
with
\begin{equation}
  {\bf S}_1\cdot{\bf T}_{24}=\frac{1}{2}(c^\dagger_{1\uparrow}c_{1\uparrow}-c^\dagger_{1\downarrow}c_{1\downarrow})
	(c_{2\uparrow}c_{4\downarrow}-c_{4\uparrow}c_{2\downarrow})+
	c^\dagger_{1\uparrow}c_{1\downarrow}c_{2\uparrow}c_{4\uparrow}+
	c^\dagger_{1\downarrow}c_{1\uparrow}c_{2\downarrow}c_{4\downarrow}.
\label{eq:16}
\end{equation}
Note that since ${\bf T}_{24}=-{\bf T}_{42}$, Eq.~(\ref{eq:15}) has
$d_{x^2-y^2}$ symmetry. Acting on the undoped groundstate, $\Delta_{\sqrt2}$
generates the diagonal singlets
\begin{equation}
  \Delta_{\sqrt2}|\psi\rangle_0\sim(\Delta^\dagger_{13}+\Delta^\dagger_{24})|0\rangle.
\label{eq:17}
\end{equation}
Thus, the projection studies of the patterns of the probable hole locations
and the exchange field $\langle{\bf S}_i\cdot{\bf S}_j\rangle$ strength show that
the stripes contain hole and spin structures which have similar features to
that of a $d$-wave pair.

\section{Conclusions}
\label{sec:4}

The domain wall structures found in doped $t$-$J$ lattices consist of relatively
narrow (1,0) stripes containing excess holes surrounded by more lightly doped
$\pi$ phase shifted antiferromagnetic regions. The linear hole density
$\rho_\ell=0.5$ along a stripe corresponds to one hole per domain wall unit
cell and the stripes repel each other. This leads to a stripe spacing
$d=(2x)^{-1}$ for a doping in which the average number of electrons per site
$\langle n_e\rangle=1-x$. For dopings larger than $x\sim0.17$, the domain
walls cease to appear as well defined structures.

In addition to the charge and spin structures which appear in the domain walls,
the application of a proximity pairfield on the boundary of a domain produces
a $d$-wave-like pairfield response. For the parameter range we have studied,
the induced pairfield extends over the system without an alternation in its
overall phase but appears to be short range. Examination of the ``most
probable" hole configurations and the exchange field $\langle{\bf S}_i\cdot{\bf S}_j\rangle$
structures associated with these configurations show patterns which reflect
the presence of $d$-wave pairing correlations in the domain walls.
Hartree-Fock calculations which neglect these pairfield correlations give
insulating domain walls with $\rho_\ell=1.0$. Thus, the pairing correlations
are an important feature of the striped phase found in the $t$-$J$ model.

DMRG calculations also find evidence for similar stripes in the Hubbard 
model \cite{ref:26,ref:27}.
In addition, dynamic cluster quantum Monte Carlo simulations of a 2D Hubbard 
model with imposed stripe-like charge-density-wave modulations find that the strength of the pairing correlations depend upon both the amplitude and wave vector of the charge modulation\cite{maier}. 
The fact that both the $t$-$J$ and Hubbard models exhibit such domain wall
structures provides support for the belief that these models contain
important parts of the basic physics associated with the high $T_c$ cuprates.

Nevertheless, there are some striking experimental results which remain to be
understood. For example \cite{ref:28}, the co-doped super-oxygenated La$_{2-x}$Sr$_x$CuO$_{4+y}$
exhibits both a long-range incommensurate antiferromagnetic phase with
$\langle n_h\rangle\simeq0.125$ and a superconducting phase with
$\langle n_h\rangle\simeq0.16$. It appears that this material phase separates
into two line compounds having different hole densities but 
with $T_N\simeq T_s\simeq40$K. In La$_2$CuO$_{4.11}$,
which has interstitial oxygen positioned in every fourth La$_2$O$_2$ layer,
spin-stripe and bulk superconductivity appear simultaneously at $T_c=42$K \cite{ref:29}.
Finally, anisotropic transport and magnetization data for
La$_{1.875}$Ba$_{0.125}$CuO$_4$ provide evidence that two-dimensional
superconducting correlations coexist with the stripe order up to temperatures
of order 40K \cite{ref:30}. The absence of 3D superconducting order above $T_c\sim4$K
suggests an antiphase ordering in the superconducting state which could then
suppress the interlayer Josephson coupling.

As noted, there are models \cite{ref:19,ref:20} that can give rise to such
intertwined phases and early variational Monte Carlo calculations \cite{ref:18}
found that for $t'<0$ stripe states with antiphase $d$-wave superconducting
order were in fact stabilized relative to stripes with in-phase superconductivity.
However, both the DMRG calculations that we have discussed and recent renormalized
mean-field theory calculations \cite{ref:16} find that the antiphase $d$-wave state
is slightly higher in energy than that of the in-phase state. Thus, while it
appears that the $t$-$t'$-$J$ models provide a good starting point for
discussing stripes, there remains more to be understood.

\section*{Acknowledgments}

DJS acknowledges the support of the Center for Nanophase Materials Science at
ORNL which is sponsored by the Division of Science User Facilities, U.S. DOE. 
SRW would like to thank the NSF for support under DMR 090-7500. 











\end{document}